\documentclass[a4paper,11pt]{article}
\pdfoutput=1 

\usepackage{jheppub} 

\usepackage[T1]{fontenc} 

\usepackage[table]{xcolor}
\definecolor{lightblue}{cmyk}{.35,0,0,0}
\definecolor{lightgreen}{cmyk}{0,0,.35,0}

\title{\boldmath [Re]constructing Finite Flavour Groups:  Horizontal Symmetry Scans from the Bottom-Up}

\author{Jim Talbert}

\affiliation{Rudolf Peierls Centre for Theoretical Physics\\University of Oxford\\1 Keble Road, OX1 3NP\\Oxford, UK}

\emailAdd{jim.talbert@physics.ox.ac.uk}

\abstract{We present a novel procedure for identifying discrete, leptonic flavour symmetries, given a class of unitary mixing matrices.  By creating explicit 3D representations for generators of residual symmetries in both the charged lepton and neutrino sector, we reconstruct large(r) non-abelian flavour groups using the GAP language for computational finite algebra.  We use experimental data to construct only those generators that yield acceptable (or preferable) mixing patterns.  Such an approach is advantageous because it 1) can reproduce known groups from other $\lq \text{top-down}$' scans while elucidating their origins from residuals, 2) find new previously unconsidered groups, and 3) serve as a powerful model building tool for theorists wishing to explore exotic flavour scenarios.  We test our procedure on a generalization of the canonical tri-bimaximal (TBM) form.}

\begin{document} 
\maketitle
\flushbottom

\section{Introduction}
\label{sec:intro}

The flavour symmetry paradigm assumes that a family symmetry $\mathcal{G_{F}}$ is broken to residual symmetries in both the charged lepton and neutrino sector:\footnote{One can also consider the breakdown of a family symmetry to both quarks and leptons, as in \cite{41}.}
\begin{equation}
\label{eq:GF}
\mathcal{G_{F}}  \rightarrow  \begin{cases}
				\mathcal{G_\nu}
				\\
				\mathcal{G_{\text{e}}}
				\end{cases}
\end{equation}
In the case of three distinguishable Majorana neutrinos, $\mathcal{G_\nu}$ can be identified with a full $Z_{2} \times Z_{2}$ Klein group or a single $Z_{2}$, with the former fully constraining the mixing matrix and the latter (generally speaking) only quantizing a single column.  $\mathcal{G_{\text{e}}}$ can be identified as a subgroup of $U(1)^{3}$, i.e. a $Z_{l}$ where $l \geqq 3$ to avoid degenerate masses for the charged leptons \cite{1}.  Both residual symmetries are abelian and finite as is consistent with non-degenerate fermion masses, whereas the parent group is generally considered to be non-abelian and finite in order to have irreducible representations (irreps) of dimension greater than 1, and thus accommodate the observed non-hierarchal mixing patterns of leptons \cite{6}.\\
\\
That residual symmetries can be used to classify full family symmetries is well known; see \cite{2,3} for recent analyses.  Furthermore, the GAP system for computational finite algebra \cite{4} has previously been used in attempts to identify viable flavour symmetries, normally by making concrete assumptions regarding the structure of the residuals and then using Lagrange's Theorem to sift through all groups in GAP's $SmallGroup$ library, up to a given order \cite{5,6,7}.  In \cite{5}, for example, two scans are performed.  The first identifies $\mathcal{G_{\nu}} = Z_{2}$ and $\mathcal{G_{\text{e}}} = Z_{2} \times Z_{2}$ and scans for parent groups up to order 1536, while the second identifies $\mathcal{G_{\nu} }= Z_{2} \times Z_{2}$ and $\mathcal{G_{\text{e}} }= Z_{l}$ and scans up to order 800.  Ref. \cite{6} only considers the maximal case $\mathcal{G_{\nu}} = Z_{2} \times Z_{2}$.  While the results of these scans vary depending on the order of parent groups considered, the assumptions regarding the structure of residuals, the origins of $\mathcal{G_{F}}$ (subgroup of SU(3) or, as in \cite{42,43}, U(3)) and types of representations they contain (scans often insist that $\mathcal{G_{F}}$ contain a faithful 3D irrep), there appears to be some consensus that at least the finite modular groups (discussed in \cite{8}) and the $\Delta \left(6 N^{2}\right)$ groups \cite{6,9,46} remain viable candidate symmetries.
\\
\\
In this paper we introduce a new method to invert the direction of the arrow in \eqref{eq:GF} by using the power of GAP to close groups generated by the $Z_{2,l}$ generators.  Our approach \textit{begins} with a class of unitary mixing matrices and subsequently finds the explicit representations of the residual generators.  These representations depend on the same parameters as the mixing matrices themselves.  After all, the matrix that diagonalizes the mass matrices of charged leptons and neutrinos also diagonalizes, up to permutations of rows and columns, the generators under which they are invariant.  By constraining these free parameters using a global fit of experimental data we are able to scan over phenomenologically viable mixing matrices,  residual generators, and finally parent symmetries.  While this method defines an arbitrarily large parameter space and hence will never be universally comprehensive, as one must always choose a finite subset of the parameter space to scan over, it can easily reproduce results from $\lq \text{top-down}$' scans and tune the assumptions made there with ease.  It can also serve as a flexible tool for model-builders who may arrive at a class of mixing matrices due to theoretical considerations made at the level of the mass matrix, of NLO corrections (e.g. charged lepton contributions), of broken symmetries, or simply pure phenomenology.
\\
\\
The paper begins in Section \ref{sec:Maths} with a discussion of the representation theory of the residual symmetries in a particularly nice basis.  We elaborate on what we mean by a "class" of mixing matrices, and the parameters upon which they depend.  We also briefly note the role of the residual generators in model-building.  The procedure used to search for flavour symmetries with GAP is outlined in Section \ref{sec:proc}.  Then the procedure is tested on a phenomenologically interesting class of matrices in Section \ref{sec:tests}.  Results of the scans and discussions are presented there as well.  
\section{Residual Symmetries and Model-Building}
\label{sec:Maths}
\subsection{The Symmetries}
The mathematical foundations of our analysis were explored by Hernandez and Smirnov in \cite{10}.  We thus largely repeat their discussion in this subsection, following also their notation.  
\\
Written in the flavour basis, the leptonic Yukawa sector Lagrangian reads:
\\
\begin{equation}
\label{eq:Lag}
\mathcal{L} = \frac{g}{\sqrt{2}} \overline{l}_{L} \gamma^{\mu} \nu_{L} W^{+}_{\mu} + \overline{E}_{R} m_{l} l_{L} + \frac{1}{2} \overline{\nu}^{c}_{L} M_{\nu U} \nu_{L} + h.c.
\end{equation}
\\
where $\nu_{L}$ and $l_{L}$ are left-handed triplets of neutrinos and charged leptons respectively, $E_{R}$ the triplet of right-handed charged leptons, and $m_{l}$ a diagonalized mass matrix of charged leptons. We have assumed that neutrinos are Majorana particles, hence the effective neutrino mass matrix presented can be seen (as per usual) as the realization of a see-saw mechanism \cite{11}.  The mismatch between flavour and mass bases is parametrized by a 3 $\times$ 3 unitary rotation matrix $U_{PMNS}$:
\\
\begin{equation}
M_{\nu U} = U^{\star}_{PMNS} m_{\nu} U^{\dagger}_{PMNS}
\end{equation}
where as always
\begin{equation}
\label{UPMNS}
U_{PMNS} = \Omega^{\dagger}_{e} \Omega_{\nu}
\end{equation}
The philosophy of \eqref{eq:GF} is that, from \eqref{eq:Lag}, one can identify the residual symmetries $\mathcal{G_{\nu}}$ and $\mathcal{G_{\text{e}}}$ by examining the invariance properties of the effective low-energy Lagrangian.  From the last term one notes that a $Z_{2} \times Z_{2}$ Klein transformation of the neutrino triplets leaves the mass matrix invariant:
\begin{equation}
\label{eq:Sgen}
\nu \rightarrow S_{i U}\nu \text{ },
\left(i = 1,2,3\right)  
\text{      and      }  S_{i U} = U_{PMNS} S_{i} U^{\dagger}_{PMNS}
\end{equation}
We work in the following diagonalized Klein basis:
\begin{equation*}
S_{1} = diag\left(1,-1,-1\right),\text{    }  S_{2} = diag\left(-1,1,-1\right),\text{    }  S_{3} = diag\left(-1,-1,1\right)
\end{equation*}
Thus we can identify $\mathcal{G_{\nu}}$ with the Klein group formed by $S_{i U}$ and $S_{j U}$, or a single $Z_{2}$ formed by $S_{i U}$.   From the charged lepton mass term, we see that there is a $U(1)^{3}$ symmetry.  We assume that $\mathcal{G_{\text{e}}} = Z_{l}$, so that it is finite.  An explicit matrix representation of the $\mathcal{G_{\text{e}}}$ is given by:
\begin{equation}
\label{eq:t}
T = diag\left(e^{i \phi_{e}}, e^{i \phi_{\mu}}, e^{i \phi_{\tau}}\right)
\text{  where  } 
\phi_{i} = 2\pi \frac{k_{i}}{l} \text{   } \text{and } i=e,\mu,\tau
\end{equation}
It is clear that the order of the generator $T$ is given by \textit{l}.  If we assume further that $\mathcal{G_{\text{e}}}$ is a subgroup of SU(3) then we can reduce the number of free charges in $T$ by 1 according to $\phi_{e} + \phi_{\mu} + \phi_{\tau} = 0$, such that:
\begin{equation}
\label{eq:SU3}
\phi_{\tau} = -2\pi \frac{k_{e}+k_{\mu}}{l}
\end{equation}
One can see that $\Omega_{e}$ in \eqref{UPMNS} reduces to the identity matrix in the basis we've chosen, as $T$ is already diagonal.  This basis is preferable because it is particularly amenable to theorists wishing to, e.g, introduce charged lepton corrections \cite{12} that may arise at a higher order, as may be motivated in analogy to the quark sector (and thus approaches related to Grand Unified Theories \cite{13,14, 44, 45}).
\\
\\
Having identified the residual symmetries and written down explicit forms for their generators, one is now in a position to $\lq \text{reconstruct}$' the parent symmetry $\mathcal{G_{F}}$, as it is merely the group of all product matrices of $S_{i U}$ and $T$.  
\subsubsection{Generators as Functions of Continuous Parameters}
The obvious (yet critical) realization for the procedure outlined in the next section is that virtually all known mixing scenarios are implementations of a wider class of matrices.  Such general classes of mixing matrices are parametrized with free variables:
\begin{equation}
U_{PMNS} \equiv U_{PMNS}\left( \lbrace \Theta_{i} \rbrace \right)
\end{equation} 
An example would be those preserving a $\mu$ - $\tau$ symmetry of the mass matrix:
\begin{equation}
\label{eq:Mt}
U_{\mu \tau}=
\left(
\begin{array}{ccc}
\cos(\theta) & \sin(\theta) & 0\\
-\frac{\sin(\theta)}{\sqrt{2}} & \frac{\cos(\theta)}{\sqrt{2}} & \frac{1}{\sqrt{2}}\\
\frac{\sin(\theta)}{\sqrt{2}} & -\frac{\cos(\theta)}{\sqrt{2}} & \frac{1}{\sqrt{2}}
\end{array}
\right)
\end{equation}
Virtually all of the canonical lowest-order forms for $U_{PMNS}$ explored before the measurement of a non-zero reactor angle \cite{15,16,17} fall within this class:
\begin{equation}
\label{eq:mt2}
U_{\mu \tau}\left(\theta\right) \rightarrow \begin{cases}
								U_{TBM}  &\rightleftarrows \tan(\theta) = \frac{1}{\sqrt{2}}
								\\
								U_{BM}  &\rightleftarrows \tan(\theta) = 1 \text{ or } \theta=\frac{\pi}{4}
								\\
								U_{GR_{i}} &\rightleftarrows \tan(\theta) = \frac{2}{(1+\sqrt{5})}  \text{, } \theta=\frac{\pi}{5}
								\end{cases}
\end{equation}
$TBM$ refers to Tri-bimaximal mixing \cite{18}, $BM$ to Bimaximal mixing \cite{19} and $GR_{i}$ to variants of Golden Ratio mixing \cite{20,21}.  Our search relies on \eqref{eq:Sgen}, which tells us that whenever  mixing matrices depend on free parameters, so do the generators $S_{i U}$:
\begin{equation}
S_{i U} \equiv S_{i U}\left( \lbrace \Theta_{i} \rbrace \right)
\end{equation}
By applying \eqref{eq:Sgen}, we find the following forms for $S^{\mu \tau}_{i U}$:
\begin{subequations}
\begin{align*}
S^{\mu \tau}_{1 U} &= 
\left(
\begin{array}{ccc}
\cos 2\theta & -\sqrt{2}\cos \theta \sin \theta & \sqrt{2} \cos \theta \sin \theta \\
-\sqrt{2} \cos \theta \sin \theta & - \cos^{2} \theta & - \sin^{2} \theta \\
\sqrt{2} \cos \theta \sin \theta & -\sin^{2} \theta & - \cos^{2} \theta \\
\end{array}
\right)\\
S^{\mu \tau}_{2 U} &= 
\left(
\begin{array}{ccc}
-\cos 2\theta & \sqrt{2}\cos \theta \sin \theta & -\sqrt{2} \cos \theta \sin \theta \\
\sqrt{2} \cos \theta \sin \theta & - \sin^{2} \theta & - \cos^{2} \theta \\
-\sqrt{2} \cos \theta \sin \theta & -\cos^{2} \theta & - \sin^{2} \theta \\
\end{array}
\right)\\
S^{\mu \tau}_{3 U} &=
\left(
\begin{array}{ccc}
-1 & 0 & 0\\
0 & 0 & 1\\
0 & 1 & 0\\
\end{array}
\right)
\end{align*}
\end{subequations}
These general forms are similar to those found in \cite{31}, with $S^{\mu \tau}_{3 U}$ the $\mu - \tau$ operator that forces the reactor angle $\theta_{13}$ to be zero and sets the atmospheric angle $\theta_{23}$ to $45^{\circ}$.  Obviously, though, the group $\mathcal{G_{F}}$ formed by all product matrices will not be finite if $\lbrace \Theta_{i} \rbrace$ are left unquantized.  Our search is built on parametrizations that discretize $\Theta_{i}$ within an experimentally determined region of the parameter space, allowing the formation of finite groups that yield matrix elements within a specified $\sigma$-range.  Such parametrizations will be outlined in the next section.
\subsection{A Note on Model-Building with $S_{iU}$}
While our procedure is model-independent, a principal attribute is the ease with which it can be used for model-building.  Most dynamical flavour models (see \cite{22,39} for reviews) introduce new scalar degrees of freedom $\phi^{\nu}$ called flavons which transform under certain irreps of $\mathcal{G_{F}}$.   However, the family symmetry is not respected by the vacua of the $\phi^{\nu}$ potentials.  The vacua must, of course, respect the residual symmetries that are already present in the effective, Majorana-enhanced SM Lagrangian.  In "direct models" (see classification between direct and indirect models presented in \cite{22}) this translates to the following invariance properties of vacuum expectation values (vevs):
\begin{equation}
\label{eq:inv}
S_{iU} \langle \phi^{\nu} \rangle = S_{jU} \langle \phi^{\nu} \rangle =  \langle \phi^{\nu} \rangle
\end{equation}
In the case where $\mathcal{G_{\nu}} = Z_{2}$, only the right equality above holds.  Hence the vevs of $\phi^{\nu}$ correspond to the invariant eigenvectors of $S_{iU}$, and therefore knowledge of the explicit representation of $S_{iU}$ is critical for the model-builder.

\section{The Procedure Outlined}
\label{sec:proc}
In this section we elaborate on how, using the representation theory just discussed, we can use GAP to search for viable flavour groups explaining a general class of mixing matrices.  
\subsection{Parameterizing the Matrix Degrees of Freedom}
The first step in the process is discretization of the parameters $\lbrace \Theta_{i} \rbrace$.  In this first attempt we do so via two rather naive parameterizations:
\begin{subequations}\label{eq:par}
\begin{align}
\tan(\Theta_{i}) &= \sqrt{\frac{b}{1-b}}\\
\Theta_{i} &= c\pi
\end{align}
\end{subequations}
where $b,c \equiv \frac{n}{m}$ and $(n,m) \in Integers$.  The first discretizes trigonometric functions of $\Theta$.  We have (without loss of generality) restricted ourselves to the unit circle, hence only the single parameter $b$ which is restricted to $b \in \left[0,1\right)$. The second is an obvious candidate for discretizing the angle $\Theta$ itself where, to avoid degeneracy, we insist that $\Theta$ lie between 0 and $2\pi$, so that $n \leq 2m$.   The correspondence between the $b/c$-parameters and \eqref{eq:mt2} is particularly simple:
\begin{subequations}
\begin{align}
\theta_{TBM} &\leftrightarrows b = \frac{1}{3}
\\
\theta_{BM} &\leftrightarrows b = \frac{1}{2} \text{ or } c=\frac{1}{4}
\\
\theta_{GR_{2}} &\leftrightarrows c = \frac{1}{5}
\end{align}
\end{subequations}
\subsection{Constraining the Parameters with Experiment}
The next step is to constrain the parameters $b$ and $c$.  We do so by directly comparing with experimental data.  Such data, though, is presented with respect to the traditional PDG parametrization of $U_{PMNS}$ \cite{24}.  In order to compare, one normally uses the standard procedure of equating unitary matrix elements:\footnote{Equating matrix elements amounts to (in most cases) equating spaces of differing degrees of freedom.  For example, $U_{PDG}$ has four degrees of freedom, whereas $U_{\mu \tau}$ only has one.  There is potentially then a region of the $\lbrace \Theta_{i} \rbrace$ space that, when maximized or minimized over the PDG parameter space, is incapable of maintaining the imposed equality.  The geometry of the hyper-spaces simply cannot intersect, and mathematics software is likely to complexify the $\theta_{i}$ being analyzed in order to increase the degrees of freedom available.  This can manifest in seemingly incoherent constraints, e.g. $\cos \left(\theta_{i}\right) > 1 \text{ or } \cos \left(\theta_{i}\right) < -1$.}
\begin{equation}
\label{eq:equate}
\parallel U^{PDG}_{ij} \parallel^{2} \text{  =  } \parallel U_{ij}\left(\lbrace \Theta_{k} \rbrace \right) \parallel ^{2} \text{ }  \equiv \text{ }  \parallel U_{ij}\left(b_{k},c_{k}\right) \parallel^{2}
\end{equation}
So, considering $\theta_{\mu \tau}$ in \eqref{eq:Mt} for illustrative purposes, equating the (1,3) elements as well as the (1,2) elements would give:
\begin{equation}
\label{eq:bmt}
\theta^{PDG}_{12} = \theta_{\mu \tau} \rightarrow b_{\mu \tau} \in [.259, .359] \mid_{3\sigma}
\end{equation}
In \eqref{eq:bmt} and in the analysis presented in the following section we use the recent experimental global fit presented in \cite{25}.  In this paper we only consider the fit for the normal hierarchy of neutrino masses.
\\ \\
However, generally speaking this procedure is too restrictive for our purposes because it immediately fixes the position of the $U_{ij}\left(\lbrace \Theta_{i} \rbrace \right)$ matrix element.  Yet the representation theory we employ can never know about the position of matrix elements;  the invariance properties of the Lagrangian under \eqref{eq:Sgen} reveal that the same matrix that diagonalizes the generators also diagonalizes the mass matrix \textit{only up to permutations of rows and columns} (see \cite{6}, e.g.).  Hence the most general statement one can make constraining the $b$ and $c$ parameters in this framework is:
\begin{equation}
\label{eq:equate2}
\parallel U^{PDG}_{min} \parallel ^{2} \text{  }  \leqq \text{  }  \parallel U_{ij}\left(b_{k},c_{k}\right) \parallel^{2} \text{  }  \leqq \text{  }  \parallel U^{PDG}_{max} \parallel ^{2}
\end{equation}
That is, any element of a matrix class can be no greater nor smaller than the largest or smallest (experimentally determined) elements of $U_{PDG}$.  This means that we can constrain \textit{matrix elements} within $x\mathcal{\sigma}$ (where x is an arbitrary integer), but we are not guaranteed to arrive at mixing angles that agree with data within $x\mathcal{\sigma}$.  This cut can easily be done at the end of the search.  Furthermore, one may impose \eqref{eq:equate2} on multiple elements of a given class.
\subsection{Constructing the Viable Generators and Closing the Groups}
Once the parameter space has been determined within a $\sigma$-range of choice one can then choose an iteration range for the integers (n,m) in \eqref{eq:par}, scanning and collecting those implementations of the two parameterizations that fall within the derived limits.  Then one can form the necessary trigonometric objects which compose mixing matrices and generators.   Due to the structure and data storage of GAP, this amounts to creating lists of the following GAP objects:
\begin{subequations}
\label{eq:GAP}
\begin{align}
\cos \left(\Theta\left(b\right)\right) &= ER\left(1-\frac{n}{m}\right)
\\
\sin \left(\Theta\left(b\right)\right) &= ER\left(\frac{n}{m}\right)
\end{align}
\text{for the first parameterization and, for the second parameterization:}
\begin{align}
\cos \left(c\right) &= \frac{E\left(2m\right)^{n} + E\left(2m\right)^{-n}}{2}
\\
\sin \left(c\right) &= \frac{E\left(2m\right)^{n} - E\left(2m\right)^{-n}}{2 E\left(4\right)}
\end{align}
\end{subequations}
$ER$ is a square root operation for a rational number $N$, $\sqrt{N}$, and $E$ returns the primitive N-th root of unity, $E\left(N\right)\equiv e^{\frac{2 \pi i}{N}}$. Once these lists are found, it is straightforward to then construct a list of the viable representations of $S_{iU}(b,c)$ by looping over combinations of \eqref{eq:GAP}.   Next, one needs to choose an iteration range for the charged lepton parameters $k_{i} \text{ and } l$ in \eqref{eq:t}.  $l$ represents the order of the generator and thus $l \geqq 3$.   For the case analyzed in this paper we choose $ -l < k_{i} <l \text{ up to} \mid l \mid = 5$.  Extending this range (or the range for (n,m)) is simply a matter of computational expense, though we find that the narrow space chosen is already rich.  Again, by looping over all possible combinations of $k_{i} \text{ and } l$ we create the generators $T_{j}$ from \eqref{eq:t}.  We also perform a memory cut on the $S_{i U}$, noting in preliminary scans that virtually all interesting results are obtained by computationally inexpensive generators (usually under 1000 bytes).  We thus remove any $S_{i U}$ consuming more than 2500 bytes.  Next, we remove any groups that quantize a null matrix element.\footnote{Numerically speaking, this cut accepts groups which quantize a squared matrix element to values greater than $10^{-6}$.}  Again, both of these cuts can be modified with ease.  For both $S_{i U} \text{ and } T_{j}$ lists we sift through the constructions and eliminate any duplicates.\\
\\
Finally, having created the unique generators $S_{i U} \text{ and } T_{j}$ in a specified interval of $\left(n,m,k_{i},l\right)$ and also a specified experimental $\sigma$-range, we are in a position to form the parent groups $\mathcal{G_{F}}$ closed by them.  GAP is capable of constructing groups directly from the matrix representations of generators.  It does so quickly using the $GroupWithGenerators$ command.  The idea is to form all groups closed  by
\begin{subequations}\
\label{subeq:groups}
\begin{align}
\label{subeq:full}
\mathcal{G_{F}} &= \lbrace S_{i U}, S_{j U}, T_{k} \rbrace
\\
\label{subeq:half}
\mathcal{G_{F}} &= \lbrace S_{i U}, T_{k} \rbrace 
\end{align}
\end{subequations}
\eqref{subeq:full} treats the case where $U_{PMNS}$ is fully constrained by $\mathcal{G_{\nu}} = Z_{2} \times Z_{2}$ \textit{and} $\mathcal{G_{F}}$ has such $\mathcal{G_{\nu}}$ as a subgroup. \eqref{subeq:half} treats the case where $U_{PMNS}$ has unquantized degrees of freedom \textit{or} where the model in consideration treats one $Z_{2}$ invariance of the mass matrix as accidental (both cases correspond to only a single equality in \eqref{eq:inv}).  This latter situation is the case, for example, in the canonical $A4$ model of Ferugglio and Altarelli \cite{23}\cite{26}.\\
\\
Before closing the viable parent groups, though, we do some filtering.  First, we test whether or not the order of $W_{i} \equiv (S_{i U} T_{k})$ is finite (and also $W_{j} \equiv S_{j U} T_{k}$ for the case of \eqref{subeq:full}), as is true whenever the parent group formed by the residuals is finite.  For those sets of generators that pass, we then test whether or not the $\mathcal{G_{F}}$ closed by them is \textit{1)} of order $\leqq$ 1000 and \textit{2)} non-abelian.  The former constraint can easily be tuned to the model-builder's preference.  We also cut those groups of order 512, as GAP's $SmallGroup$ library does not assign a unique ID for them.  Whenever a group is formed, we collect the associated parameters $(b, c)$ and the explicit form of $T$ (for generator and mixing matrix reconstruction).\\
\\
Finally, then, we have created/found the non-abelian groups of order $\leqq$ 1000 (excluding groups of order 512) closed by the "phenomenologically viable" generators in \eqref{subeq:groups}, within a pre-selected iteration range for the variables $\left(n,m,k_{i},l\right)$ and an experimentally determined $\sigma$-range.  Having done so, we identify the $GroupID$ and $StructureDescription$ of the group and couple this information to the associated group parameters.  
\subsection{Summary of the Steps}
\begin{flushleft}
Before executing the algorithm on an interesting class of matrices in the following section, we summarize the procedure:
\end{flushleft}
\begin{enumerate}
\item Discretize all degrees of freedom present in the class of matrices under consideration via \eqref{eq:par}
\item Constrain those parameters via \eqref{eq:equate2}
\item Construct the algebraic and/or trigonometric objects necessary to fully construct the mixing matrix, and thus the generators $S_{i U}$ via \eqref{eq:GAP}
\item Form the explicit representations of phenomenologically viable $S_{i U} \text{, and } T_{k}$ via \eqref{eq:Sgen} and \eqref{eq:t}
\item Form all finite, non-abelian groups closed by either 2 (one $Z^{\nu}_{2}$) or 3 (full $Z^{\nu}_{2} \times Z^{\nu}_{2}$) generators
\item Analyze 
\end{enumerate}

\section{An Interesting Case Study}
\label{sec:tests}
Having developed a program for symmetry searching directly from a class of mixing matrices, we now execute said algorithm on a particularly interesting case:  a generalization of the canonical TBM form \eqref{eq:mt2}.
\subsection{A Perturbation to  Tri-bimaximal Mixing}
\eqref{eq:Mt} cannot be experimentally viable without additional considerations; the reactor angle $\theta_{13}$ is vanishing.  However, many models still consider the TBM matrix \eqref{eq:mt2} a lowest order form that, due to a variety of possible corrections, becomes viable.  Such corrections could include a non-diagonal charged lepton mixing matrix $\Omega_{e}$ \cite{27,28}, an additional neutrino species in extra dimensions \cite{29}, or a hybrid mass generation mechanism \cite{30}.   In this section we look at the case where the TBM matrix is modified by a rotation in the $\left(1, 3\right)$ sector:
\begin{equation}
\label{eq:theta13}
U_{TBM}^{13} \equiv
\left(
\begin{array}{ccc}
\sqrt{\frac{2}{3}}\cos{\psi} & \frac{1}{\sqrt{3}} & \sqrt{\frac{2}{3}}\text{ }   \sin{\psi}\text{ }  e^{-i \phi}\\
- \frac{\cos{\psi}}{\sqrt{6}} - \frac{e^{i \phi} \text{ } \sin{\psi}}{\sqrt{2}} & \frac{1}{\sqrt{3}} & \frac{\cos{\psi}}{\sqrt{2}} - \frac{e^{-i \phi} \text{ } \sin{\psi}}{\sqrt{6}}\\
\frac{\cos{\psi}}{\sqrt{6}} - \frac{e^{i \phi} \text{ } \sin{\psi}}{\sqrt{2}} & -\frac{1}{\sqrt{3}}& \frac{\cos{\psi}}{\sqrt{2}} + \frac{e^{-i \phi} \text{ } \sin{\psi}}{\sqrt{6}}
\end{array}
\right)
\end{equation}
The (1,3) rotation can be motivated by considering soft symmetry breaking effects at the level of the Lagrangian \cite{31} and/or by employing additional flavons when model building \cite{32}.  At the moment these model-dependent considerations are irrelevant for our purposes.  \eqref{eq:theta13} is seen simply as a class of matrices with a phase $\left(\phi \right)$ and rotational $\left(\psi \right)$ degree of freedom.  Note that the second column remains unchanged by this rotation.  Applying \eqref{eq:Sgen} gives the following $Z_{2}$ generators:
\begin{subequations}
\begin{align*}
S^{13}_{1 U} &= \frac{1}{3}
\left(
\begin{array}{ccc}
 \left(-1 + 2 c_{2 \psi}\right)& 2 c_{\psi} \left(-c_{\psi} - \sqrt{3} \text{ } e^{-i \phi} \text{ } s_{\psi} \right) & 2 c_{\psi} \left( c_{\psi} - \sqrt{3} \text{ } e^{-i \phi} \text{ } s_{\psi} \right)\\
-2 c_{\psi} \left( c_{\psi} + \sqrt{3} \text{ } e^{i \phi} \text{ } s_{\psi} \right) & -2 c_{\psi} \left( c_{\psi} - \sqrt{3}\text{ } c_{\phi} \text{ } s_{\psi} \right) & \left(1 - 2 c_{2 \psi} - i \sqrt{3}\text{ } s_{\phi} \text{ }s_{ 2 \psi} \right) \\
2 c_{\psi} \left(c_{\psi} - \sqrt{3} \text{ } e^{i \phi} \text{ } s_{\psi} \right) &  \left(1 - 2 c_{2 \psi} + i \sqrt{3}\text{ } s_{\phi} \text{ } s_{ 2 \psi} \right) & -2 c_{\psi} \left( c_{\psi} + \sqrt{3}\text{ } c_{\phi} \text{ } s_{\psi} \right) \\
\end{array}
\right)\\
S^{13}_{2 U} &= \frac{1}{3}
\left(
\begin{array}{ccc}
-1 & 2 & -2 \\
2 & -1 & -2 \\
-2 & -2 & -1 \\
\end{array}
\right)\\
S^{13}_{3 U} &= \frac{1}{3}
\left(
\begin{array}{ccc}
 \left(-1 - 2 c_{2 \psi}\right)& 2 s_{\psi} \left(-s_{\psi} + \sqrt{3} \text{ } e^{-i \phi} \text{ } c_{\psi} \right) & 2 s_{\psi} \left( s_{\psi} + \sqrt{3} \text{ } e^{-i \phi} \text{ } c_{\psi} \right)\\
2 s_{\psi} \left( -s_{\psi} + \sqrt{3} \text{ } e^{i \phi} \text{ } c_{\psi} \right) & -2 s_{\psi} \left(s_{\psi} + \sqrt{3}\text{ } c_{\phi} \text{ } c_{\psi} \right) &  \left(1 + 2 c_{2 \psi} + i \sqrt{3}\text{ } s_{\phi} \text{ } s_{ 2 \psi} \right) \\
2 s_{\psi} \left( s_{\psi} + \sqrt{3} \text{ } e^{i \phi} \text{ } c_{\psi} \right) &  \left(1 + 2 c_{2 \psi} - i \sqrt{3}\text{ } s_{\phi} \text{ } s_{ 2 \psi} \right) & 2 s_{\psi} \left( -s_{\psi} + \sqrt{3}\text{ } c_{\phi} \text{ } c_{\psi} \right) \\
\end{array}
\right)
\end{align*}
\end{subequations}
where $s_{\psi}$ and $c_{\psi}$ stand for $\sin\psi$ and $\cos\psi$ respectively.  Note that $S^{13}_{2U} = S^{\mu\tau}_{2U}\vert_{TBM}$, a fact correlated to the unchanged second column in $U^{13}_{TBM}$.  The other two matrices are composed of simple trigonometric functions of $\psi \text{ and } \phi$, hence the GAP objects in \eqref{eq:GAP} are what we construct.  As always, the charged lepton generator is given by \eqref{eq:t}, while in this scan we also apply the SU(3) constraint \eqref{eq:SU3}.  Furthermore, motivated by \cite{6} and the fact that $\phi$ is poorly constrained because of its dependence on $\theta_{23}$ (apply \eqref{eq:equate} to see this), we simplify the structure of \eqref{eq:theta13} by preselecting interesting values of the phase $\phi$.  $\phi$ is related to the physical CP violating phase $\delta$ via the following equation:
\begin{equation}
\label{eq:delta}
\cos{\delta} = \frac{2 \cos{\phi} \left(1 + 2 \cos{2 \psi}\right)}{\sqrt{15 + 16 \cos{2 \psi} + 5 \cos{4 \psi} - 6 \cos{2 \phi} \left(\sin{2 \psi}\right)^{2}}}
\end{equation} 
Hence corresponding values of $\delta$ are: 
\begin{equation}
\label{eq:delta2}
\cos{\delta} =						 \begin{cases}
								1 &\rightleftarrows \phi = 0
								\\
								0  &\rightleftarrows \phi = \frac{\pi}{2}
								\\
								\frac{1 + 2 \cos{2 \psi}}{\sqrt{8 \cos{2 \psi} + \frac{5}{2}\left(3 + \cos{4 \psi}\right)}} &\rightleftarrows \phi = \frac{\pi}{4}
								\end{cases}
\end{equation}
 Clearly $\phi = \frac{\pi}{4}$  directly couples the CP violating phase to $\psi$, and so in this case $\delta$ can take a range of values between 0 and 2$\pi$.\footnote{Note that in this study we only apply the second discretization scheme in \eqref{eq:par} to $\psi$, for the case $\phi = \frac{\pi}{4}$.}  By making the choices in \eqref{eq:delta2}, $U^{13}_{TBM}$ only depends on one degree of freedom.  Recalling \eqref{eq:equate2} and using only the $\left(1,3\right)$ element of $U^{13}_{TBM}$ we find (within 3$\sigma$ of the PDG elements):
 \begin{equation}
 .0176 \text{ } \leqq \text{ } \frac{2}{3} (\sin{\psi})^{2} \text{ } \leqq \text{ } .728 \text{ } \footnote{This upper bound constitutes a unitarity violation in the context of $U^{13}_{TBM}$, hence one could effectively reduce this  to .667.  While this is not done at the level of the code, the results in Appendix \ref{sec:tables} show that no quantized (squared) matrix element lies outside of this effective limit.  Solving the right equality, one also sees an example of the situation discussed in the first footnote, where $\sin \left(\psi\right) > 1$, suggesting that the limits of validity of \eqref{eq:equate} correspond to unitarity limits.}
 \end{equation}
Because $\psi$ is the only degree of freedom, we only have to choose one scan range for the variables $\left(n,m\right)$.  We choose $\left(n,m\right) \in \left[0/1..20/21\right]$ (where the slash differentiates between the first and second discretization schemes \eqref{eq:par} respectively), and again $l \in \left[3..5\right]$ with $-5 < k_{i} <5$.  The results of these scans are found in Appendix \ref{sec:tables}.  
\subsection{Discussion}
The results presented in Tables \ref{tab:one}, \ref{tab:two}, and \ref{tab:three} in Appendix \ref{sec:tables} are rich, especially considering the limited parameter space scanned in this first attempt.  While many groups are found, only 5 are capable of unambiguously quantizing mixing angles within $3\sigma$ of their global fit (see Appendix \ref{sec:tables} for details on group structure):   $\Delta(600)$, $\Delta(150)$, $Z_{3} \times \Delta(150)$, $\Delta(726)$, and $\Xi(18,6)$.  Of these, only $\Delta(600)$ and $\Xi(18,6)$ impose the full Klein symmetry of the Majorana mass matrix and thus quantize all 3 mixing angles.  The values of these angles are found in Table \ref{tab:angles}.  The ambiguity in the prediction for $\theta_{23}$ is due to the permutation freedom of the rows mentioned in Section \ref{sec:proc}.  In all 5 physically promising cases the symmetries predict a vanishing physical CP violating phase $\delta$.  These results agree exactly with the results found in \cite{6}.  We have also found that $\Delta(150)$ is the smallest group capable of successfully quantizing the third column of $U^{13}_{TBM}$ and hence the reactor angle $\theta_{13}$, a result that fully agrees with \cite{7} and \cite{41}.  Furthermore, the column(s) quantized by $\Delta(96)$ and $\Delta(384)$ are precisely those found in \cite{10}.  It is clear then that our algorithm produces results that are beautifully consistent with former approaches, yet with the added benefits of bottom-up $\lq \text{[re]construction.}$' Interestingly, none of the groups predicting non-vanishing CP violation are consistent with experiment.  That is, we only found groups that predict vanishing CP violating phase $\delta$ yield mixing angles within 3$\sigma$ of their global fits. 
\begin{table}
\centering
\begin{tabular}{|c|c|c|c|c|}
\hline \hline
Group & $\sin^{2}\theta_{12}$& $\sin^{2}\theta_{13}$ & $\sin^{2}\theta_{23}$ & $\cos \delta$\\
\hline \hline
$\Delta(600)$ & .3432 & .0288 & .6209 or .3791 & 1\\
\hline
$\Xi(18,6)$ & .3402 & .0201 & .6008 or .3992 & 1\\
\hline\hline
\end{tabular}
\caption{Quantized Mixing Angles}
\label{tab:angles}
\end{table}
Note also that in all Tables in Appendix \ref{sec:tables} the tetrahedral group $A_{4}$ is found for the case where $\mathcal{G_{F}} = \lbrace S^{13}_{2 U}, T \rbrace$.  This is completely unsurprising as $S^{13}_{2 U}$ is the generator associated with the invariance of the second column of $U^{13}_{TBM}$, which is equivalent to the second column and $\lq$second' generator of $U_{TBM}$.  Indeed, the generators of $A_{4}$ are presented in a similar basis in \cite{23} where Ferugglio and Altarelli derive TBM mixing from the tetrahedral group structure.  Had we studied the class of matrices where $U_{TBM}$ is modified by a rotation in the (2, 3) sector as opposed to the (1, 3) sector, the first column would be unmodified from its original TBM form.  The associated neutrino generator, when combined with a $Z_{3}$ charged lepton generator, would instead close the cubic group $S_{4}$.\\  
\\
We have also found groups that, while not immediately viable from symmetry considerations alone, may become so after NLO corrections.  As an exotic example we note a recent proposal \cite{35} where the reactor angle is \textit{reduced} by charged lepton corrections (as opposed to augmented from 0), perhaps on the order of the Wolfenstein parameter \cite{24, 36}.  In such a scenario the reactor angle would originally be quantized at $\theta_{13} \sim 18^{\circ}$.  Intriguingly, our search reveals that the groups $\Xi(9,3)$ and $\Xi(18,6)$, $\Delta(384)$, and $Z_{5} \times D_{10}$ and $\Sigma(32)$  can yield $\theta_{13} \simeq 16.2^{\circ}, \text{ } 18.2^{\circ}, \text{ and } 15.7^{\circ}$ respectively.  While we do not explore this issue further here, the point is that our method can give useful information to model-builders who may be wishing to justify purely phenomenological considerations such as those in \cite{35}. 
\section{Conclusions}
 \label{sec:Conc}
Despite being somewhat underdetermined, the flavour symmetry paradigm could potentially provide an elegant and powerful solution to (part of) the Flavour Problem.  Discrete flavour symmetries imposed via finite groups are not only phenomenologically valid, they can also be extremely well motivated from high-energy theories at the GUT scale (see \cite{34} for a recent analysis of discrete symmetries in the context of F-Theory GUT models) or even beyond (see \cite{37}\cite{38}\cite{40} for studies of finite groups from extra-dimensional orbifolds, both with and without string selection rules).\\
\\
In this note we have introduced a novel method for identifying flavour symmetries capable of postdicting the parameters of the PMNS neutrino mixing matrix, including the currently unknown CP violating phase $\delta$.  Ours is a $\lq$bottom-up' approach as we begin with a class of mixing matrices, identify the generators of the residual symmetries present in the neutrino and charged lepton sectors, and then implement a discretization scheme to close finite groups directly constrained by experimental data with the GAP system for computational algebra.  As has been shown, various theoretical and phenomenological considerations yield different classes of matrices such as $U_{\mu \tau} \text{ or } U^{13}_{TBM}$.  We have tested our algorithm on the latter, a promising generalization of TBM mixing, and find numerous  groups.  Of these, only $\Delta(600)$ and $\Xi(18,6)$ can quantize all three mixing angles within $3\sigma$ of a current global fit of data (given the parameter space studied).  In both instances $\delta$ is predicted to vanish.  Our results appear fully consistent with numerous former approaches to model-independent studies of leptonic flavour symmetries.\\  
\\
Yet this program is still in early stages.  Future work will not only see the application of the algorithm to new and broader classes of matrices such as those presented in \cite{32}, but also to increased computational efficiency.  Improved and/or additional discretization schemes to those presented in \eqref{eq:par} should also be explored.

\appendix
\section{Tables of Results}
\label{sec:tables}
This appendix lists the results of our scans as described in Section \ref{sec:tests}.  The first column gives the neutrino generator(s) $S_{i U}\left(b,c\right)$ while the second column gives the explicit form of the charged lepton generator $T$ ($\omega \equiv e^{\frac{i 2 \pi}{3}},\text{ } \rho \equiv e^{\frac{i 2 \pi}{4}},\text{ } \lambda \equiv e^{\frac{i 2 \pi}{5}}$). That is, if $(1)$ is in the first column then the group listed is closed by $\lbrace S_{1 U}, T \rbrace$, while if $(12)$ is in the first column the group is closed by $\lbrace S_{1 U}, S_{2 U}, T \rbrace$.   The asterisk $(*)$, degree $(\circ)$, and dagger $(\dagger)$ symbols indicate that the information is also relevant for the $S_{i U}$ tagged.  The $(\circ)$ symbol further implies a swap in values between the first and third columns, which can be calculated for $U^{13}_{TBM}$ via unitarity.  In cases where more than one value of $(b,c)$ yields the same group (and the same quantized eigenvector, up to permutations of the elements) or when different forms of $T$ give the same group, then only one value or form is presented.\\  
\\
The third column gives the explicit value of $(b,c)$ used, where double horizontal lines differentiate between the first and second schemes (i.e. b and c).  The fourth and fifth columns give the $SmallGroup$ ID assigned by the GAP system and the $StructureDescription$, or common name of the group.  Our naming scheme follows the comprehensive review found in \cite{33} when available, while we have named the $\Xi(N,M)$ groups ourselves for simplicity:
\begin{subequations}
\begin{align}
\Delta(6N^{2}) &\equiv \left( \left( Z_{N} \times Z_{N} \right) \rtimes Z_{3} \right) \rtimes Z_{2} 
\\
\Sigma(2N^{2}) &\equiv \left( Z_{N} \times Z_{N} \right) \rtimes Z_{2}
\\
\Xi(N,M) &\equiv \left( \left( Z_{N} \times Z_{M} \right) \rtimes Z_{3} \right) \rtimes Z_{2}
\end{align}
\end{subequations}
The sixth column gives the squared elements of the third column of the mixing matrix:  $\vert\vert U_{i3}^{T} \vert\vert^{2} = (\vert U_{13} \vert^{2}, \vert U_{23} \vert^{2}, \vert U_{33} \vert^{2})$.  For groups listed in the $(1)$ row, the actual eigenvector quantized corresponds to the $\textit{first}$ column of $U^{13}_{TBM}$.  To illustrate all of the above, the row of parameters $\parallel (1),(3)^{\circ} \parallel \frac{1}{14}, \frac{3}{7}^{\circ} \parallel \text{ }[.0330, .358, .609]^{\circ} \parallel$ indicates that $\mathcal{G_{F}} = \lbrace S_{1U}(c=\frac{1}{14}), T \rbrace$  and $\mathcal{G_{F}} = \lbrace S_{3U}(c=\frac{3}{7}), T \rbrace$ have invariant eigenvectors with squared moduli [.634, .308, .0579], up to permutations.  This vector can correspond to any column of the physical mixing matrix.  For the case of $\phi=\frac{\pi}{4}$, the last column gives the values of $\cos \delta$.  Finally, highlighted cells indicate that the relevant mixing angles are accommodated within $3\sigma$ of their experimentally measured values as presented in the global fit \cite{25}.  Blue indicates that the full Klein symmetry is imposed and thus all 3 mixing angles are accommodated whereas yellow indicates that only one $Z_{2}$ neutrino generator is present.
\begin{table}
\label{tab:one}
\makebox[\textwidth]{
\begin{tabular}{|c|c|c|c|c|c|}
\hline \hline
(i,j) in $\lbrace S_{i U}, S_{j U} \rbrace$ & $T_{diag}$ & b or c & GAP-ID & Group Structure & $\parallel U_{i 3}^{2} \parallel ^{T}$ \\
\hline \hline
(1), $(3)^{*}$ & [$\omega^{2}$, 1, $\omega$] & $\frac{1}{2}, \frac{1}{2}^{*\dagger}$ & [288, 397] & $Z_{3} \times \Delta(96)$ & $[.333, .0447, .622]^{*\dagger}$\\
$(12, 13, 23)^{\dagger}$  & [1, $\omega^{2}$, $\omega$] & $\frac{1}{2}, \frac{1}{2}^{*\dagger}$ & [96, 64] & $\Delta(96)$ & $[.333, .0447, .622]^{*\dagger}$\\
\hline
(2) & [$\omega^{2}$, 1, $\omega$] & N.A. & [12, 3] & $A_{4}$ & N.A.\\
\hline \hline
(1) &  [$\omega^{2}$, 1, $\omega$] & $\frac{1}{4}, \frac{1}{4}^{*\dagger}$ & [288, 397] & $Z_{3} \times \Delta(96)$ & $[.333, .0447, .622]^{*\dagger}$\\
$(3)^{*}$, $(3)^{\circ}$ & [1, $\omega^{2}$, $\omega$]& $\frac{1}{4}, \frac{1}{4}^{*\dagger}$ & [96, 64] & $\Delta(96)$ & $[.333, .0447, .622]^{*\dagger}$\\
$(12, 13, 23)^{\dagger}$ & [1, $\omega^{2}$, $\omega$] & $\frac{1}{5}, \frac{3}{10}^{\circ}, \frac{1}{5}^{\dagger}$ & [600, 179] & $\Delta(600)$ & $[.230, .110, .659]^{\circ\dagger}$\\
&  [1, $\omega^{2}$, $\omega$] & $\frac{1}{8}, \frac{1}{8}^{*\dagger}$ & [384, 568] & $\Delta(384)$ & $[.0976, .247, .655]^{*\dagger}$\\
&  [1, $\omega^{2}$, $\omega$] & $\frac{3}{8}, \frac{3}{8}^{*\dagger}$ & [384, 568] & $\Delta(384)$ & $[.569, .0114, .420]^{*\dagger}$\\
&  [$\omega^{2}$, 1, $\omega$] & $\frac{1}{9}, \frac{1}{18}^{\circ}, \frac{1}{9}^{\dagger}$ & [648, 259] & $\Xi(18, 6)$ & \cellcolor{lightblue}$[.0780, .276, .647]^{\circ\dagger}$\\
&  [$\omega^{2}$, 1, $\omega$] & $\frac{1}{10}, \frac{2}{5}^{\circ}$ & [450, 20] & $Z_{3} \times \Delta(150)$ & \cellcolor{lightgreen}$[.0637, .299, .638]^{\circ}$\\
&  [1, $\omega^{2}$, $\omega$] & $\frac{1}{10}, \frac{2}{5}^{\circ}$ & [150, 5] & $\Delta(150)$ & \cellcolor{lightgreen}$[.0637, .299, .638]^{\circ}$\\
&  [$\omega^{2}$, 1, $\omega$]  & $\frac{1}{14}, \frac{3}{7}^{\circ}$ & [882, 38] & $Z_{3} \times \Delta(294)$ & $[.0330, .358, .609]^{\circ}$\\
&   [1, $\omega^{2}$, $\omega$]  & $\frac{1}{14}, \frac{3}{7}^{\circ}$ & [294, 7] & $ \Delta(294)$ &  $[.0330, .358, .609]^{\circ}$\\
&   [1, $\omega^{2}$, $\omega$]  & $\frac{2}{5}, \frac{1}{10}^{\circ\dagger}$ & [600, 179] & $ \Delta(600)$ & \cellcolor{lightblue}$[.0288, .368, .603]^{\circ\dagger}$\\
&   [$\omega^{2}$, 1, $\omega$]  & $\frac{1}{18}, \frac{1}{9}^{\circ}$ & [162, 14] & $ \Xi(9, 3)$ & $[.391, .0201, .589]^{\circ}$\\
&   [$\omega^{2}$, 1, $\omega$]  & $\frac{3}{10}, \frac{1}{5}^{\circ}$ & [450, 20] & $Z_{3} \times \Delta(150)$ & $[.436, .00728, .556]^{\circ}$\\
&   [1, $\omega^{2}$, $\omega$]  & $\frac{3}{10}, \frac{1}{5}^{\circ}$ & [150, 5] & $ \Delta(150)$ & $[.436, .00728, .556]^{\circ}$\\
&  [$\omega^{2}$, 1, $\omega$]  & $\frac{5}{14}, \frac{1}{7}^{\circ}$ & [882, 38] & $Z_{3} \times \Delta(294)$ & $[.541, .00372, .455]^{\circ}$\\
&   [1, $\omega^{2}$, $\omega$]  & $\frac{5}{14}, \frac{1}{7}^{\circ}$ & [294, 7] & $ \Delta(294)$ & $[.541, .00372, .455]^{\circ}$\\
&  [$\omega^{2}$, 1, $\omega$]  & $\frac{3}{14}, \frac{2}{7}^{\circ}$ & [882, 38] & $Z_{3} \times \Delta(294)$ & $[.259, .0890, .652]^{\circ}$\\
&   [1, $\omega^{2}$, $\omega$]  & $\frac{3}{14}, \frac{2}{7}^{\circ}$ & [294, 7] & $\Delta(294)$ & $[.259, .0890, .652]^{\circ}$\\
\hline
(2) & [$\omega^{2}$, 1, $\omega$] & N.A. & [12, 3] & $A_{4}$ & N.A.\\
\hline
(3) & [1, $\omega^{2}$, $\omega$] & $\frac{1}{11}$ & [726,5] & $\Delta(726)$ & [.0529, .318, .630]\\
& [1, $\omega^{2}$, $\omega$] & $\frac{2}{11}$ & [726,5] & $\Delta(726)$ & [.195, .665, .140]\\
& [1, $\omega^{2}$, $\omega$] & $\frac{3}{11}$ & [726,5] & $\Delta(726)$ & \cellcolor{lightgreen} [.381, .0239, .595]\\
& [1, $\omega^{2}$, $\omega$] & $\frac{4}{11}$ & [726,5] & $\Delta(726)$ & [.552, .00602, .442]\\
& [1, $\omega^{2}$, $\omega$] & $\frac{5}{11}$ & [726,5] & $\Delta(726)$ & [.653, .0921, .255]\\
\hline
\end{tabular}}
\caption{Flavour Symmetries of $U^{13}_{TBM}$ ($\phi = 0$, $\cos \delta = 1$)}
\label{tab:one}
\end{table}
\begin{table}[tp]
\label{tab:two}
\makebox[\textwidth]{
\begin{tabular}{|c|c|c|c|c|c|}
\hline \hline
(i,j) in $\lbrace S_{i U}, S_{j U} \rbrace$ & $T_{diag}$ & b or c & GAP-ID & Group Structure & $\parallel U_{i 3}^{2} \parallel ^{T}$ \\
\hline \hline
(1) & [$\omega^{2}$, 1, $\omega$] & $\frac{1}{2}, \frac{1}{2}^{*\dagger}$ & [12, 3] & $A_{4}$ & $[.333, .333, .333]^{*\dagger}$\\
$(3)^{*}$, $(3)^{\circ}$ & [1, $\rho$, -$\rho$] & $\frac{1}{4}, \frac{3}{4}^{\circ}$ & [24, 12] & $S_{4}$ & $[.167, .417, .417]^{\circ}$\\
$(12,13, 23)^{\dagger}$ & [$\lambda$, $\lambda^{2}$, $\lambda^{2}$] & $\frac{1}{4}, \frac{3}{4}^{\circ}$ & [50, 3] & $Z_{5} \times D_{10}$ & $[.167, .417, .417]^{\circ}$\\
& [-1, $\rho$, $\rho$] & $\frac{1}{4}, \frac{3}{4}^{\circ}$ & [32, 11] & $\Sigma(32)$ & $[.167, .417, .417]^{\circ}$\\
\hline
(2) & [$\omega^{2}$, 1, $\omega$] & N.A. & [12, 3] & $A_{4}$ & N.A.\\
\hline \hline
(1) & [$\omega^{2}$, 1, $\omega$] & $\frac{1}{4}, \frac{1}{4}^{*\dagger}$ & [12, 3] & $A_{4}$ & $[.333, .333, .333]^{*\dagger}$\\
$(3)^{*}$, $(3)^{\circ}$ & [$\lambda$, $\lambda^{2}$, $\lambda^{2}$] & $\frac{1}{6}, \frac{1}{3}^{\circ}$ & [50, 3] & $Z_{5} \times D_{10}$ & $[.167, .417, .417]^{\circ}$\\
$(12,13, 23)^{\dagger}$& [1, $\rho$, -$\rho$] & $\frac{1}{6}, \frac{1}{3}^{\circ}$ & [24, 12] & $S_{4}$ & $[.167, .417, .417]^{\circ}$\\
& [-1, $\rho$, $\rho$]& $\frac{1}{6}, \frac{1}{3}^{\circ}$ & [32, 11] & $\Sigma(32)$ & $[.167, .417, .417]^{\circ}$\\
\hline
(2) & [$\omega^{2}$, 1, $\omega$] & N.A. & [12, 3] & $A_{4}$ & N.A.\\
\hline
\end{tabular}}
\caption{Flavour Symmetries of $U^{13}_{TBM}$ ($\phi = \frac{\pi}{2}$, $\cos \delta = 0$)}
\label{tab:two}
\end{table}

\begin{table}[tp]
\label{tab:three}
\makebox[\textwidth]{
\begin{tabular}{|c|c|c|c|c|c|c|}
\hline \hline
(i,j) in $\lbrace S_{i U}, S_{j U} \rbrace$ & $T_{diag}$ & c & GAP-ID & Group Structure & $\parallel U_{i 3}^{2} \parallel ^{T}$ & $\cos \delta$ \\
\hline \hline
(1), $(3)^{\circ}$ & [$\lambda$, $\lambda^{2}$, $\lambda^{2}$]  & $\frac{1}{6}, \frac{1}{3}^{\circ}$ & [50, 3] & $Z_{5} \times D_{10}$ & $[.167, .240, .593]^{\circ}$ & $.700, .678^{\circ}$ \\
 & [-1, $\rho$, $\rho$] & $\frac{1}{6}, \frac{1}{3}^{\circ}$ & [32, 11] &$\Sigma(32)$ & $[.167, .240, .593]^{\circ}$ & $.700, .678^{\circ}$ \\
& [$\lambda$, $\lambda^{2}$, $\lambda^{2}$] & $\frac{5}{6}, \frac{2}{3}^{\circ}$ & [50, 3] & $Z_{5} \times D_{10}$ & $[.167, .240, .593]^{\circ}$ & $.391, .550^{\circ}$\\
& [-1, $\rho$, $\rho$] & $\frac{5}{6}, \frac{2}{3}^{\circ}$ & [32, 11] & $\Sigma(32)$ & $[.167, .240, .593]^{\circ}$ & $.391, .550^{\circ}$\\
& [$\lambda$, $\lambda^{2}$, $\lambda^{2}$] & $\frac{7}{6}, \frac{4}{3}^{\circ}$ & [50, 3] & $Z_{5} \times D_{10}$ & $[.167, .240, .593]^{\circ}$ & $-.280, -.574^{\circ}$\\
& [-1, $\rho$, $\rho$] & $\frac{7}{6}, \frac{4}{3}^{\circ}$ & [32, 11] & $\Sigma(32)$ & $[.167, .240, .593]^{\circ}$ & $-.280, -.574^{\circ}$\\
& [$\lambda$, $\lambda^{2}$, $\lambda^{2}$] & $\frac{11}{6}, \frac{5}{3}^{\circ}$ & [50, 3] & $Z_{5} \times D_{10}$ & $[.167, .240, .593]^{\circ}$ & $-.541, -.687^{\circ}$\\
& [-1, $\rho$, $\rho$] & $\frac{11}{6}, \frac{5}{3}^{\circ}$ & [32, 11] & $\Sigma(32)$ & $[.167, .240, .593]^{\circ}$ & $-.541, -.687^{\circ}$\\
\hline
(2) & [$\omega^{2}$, 1, $\omega$] & N.A. & [12, 3] & $A_{4}$ & N.A. & N.A.\\
\hline
\end{tabular}}
\caption{Flavour Symmetries of $U^{13}_{TBM}$ ($\phi = \frac{\pi}{4}$, $\cos \delta = \frac{1 + 2 \cos{2 \psi}}{\sqrt{8 \cos{2 \psi} + \frac{5}{2}\left(3 + \cos{4 \psi}\right)}}$)}
\label{tab:three}
\end{table}

\clearpage
\acknowledgments
I am most grateful to Professor Graham Ross for many helpful discussions, guiding insights and critical suggestions throughout the duration of this project.  I am also grateful to J\"{u}rgen Rohrwild for discussions regarding the footnote on page (6) and comments on the manuscript, to Rasmus Rasmussen for helpful literature reviews, and to Jonathan Patterson for assisting with the Hydra computing cluster which ran the scans.  I also acknowledge the University of Oxford Department of Physics and the Niels Bohr Institute for travel support during the course of this research.




\end{document}